\documentclass[12pt,preprint]{aastex}

\shorttitle{RRLs with the EVLA}
\shortauthors{Kepley et al.}

\newcommand{\ha}{\ensuremath{{\rm H}\alpha}}
\newcommand{\kms}{\ensuremath{\rm{km \, s}^{-1}}}

\newcommand{\Jybeam}{\ensuremath{\rm{Jy \ beam^{-1}} \ }}

\newcommand{\mJybeam}{\ensuremath{\rm{mJy \ beam^{-1}} \ }}

\newcommand{\hrrl}[1]{H{#1}$\alpha$}

\begin{document}

%%----------------------------------------------------------------------
%                          Front Matter
%%----------------------------------------------------------------------

\title{Unveiling Extragalactic Star Formation Using Radio
Recombination Lines: An EVLA Pilot Study with NGC 253}

\author{Amanda A. Kepley\altaffilmark{1}}
\affil{Department of Astronomy, University of Virginia, P.O. Box 400325, Charlottesville, VA 22904-4325}
\email{kepley@virginia.edu}

\author{Laura Chomiuk\altaffilmark{2}}
\affil{National Radio Astronomy Observatory, 520 Edgemont Road, Charlottesville, VA 22903-2475}
\affil{Harvard-Smithsonian Center for Astrophysics, 60 Garden Street, Cambridge, MA 02138}
\email{lchomiuk@cfa.harvard.edu}

\author{Kelsey E. Johnson\altaffilmark{3}}
\affil{Department of Astronomy, University of Virginia, P.O. Box 400325, Charlottesville, VA 22904-4325}
\email{kej7a@virginia.edu}

\author{W.M. Goss}
\affil{National Radio Astronomy Observatory, Pete V. Domenici Science Operations Center, P.O. Box 0, 1003 Lopezville Road, Socorro, NM 87801-0387 }
\email{mgoss@nrao.edu}

\author{Dana S. Balser} \affil{National Radio Astronomy Observatory,
520 Edgemont Road, Charlottesville, VA 22903-2475 }
\email{dbalser@nrao.edu}

\and

\author{D.J. Pisano\altaffilmark{3}}
\affil{Department of Physics, West Virginia University, 244 Hodges Hall, P.O. Box 6315, Morgantown, WV 26506}
\email{DJPisano@mail.wvu.edu}

\altaffiltext{1}{Visiting Research Associate at the National Radio Astronomy Observatory}

\altaffiltext{2}{Jansky Fellow of the National Radio Astronomy Observatory.}

\altaffiltext{3}{Adjunct Assistant Astronomer at the National Radio
Astronomy Observatory}

\begin{abstract}

  Radio recombination lines (RRLs) are powerful, extinction-free
  diagnostics of the ionized gas in young, star-forming
  regions. Unfortunately, these lines are difficult to detect in
  external galaxies.  We present the results of EVLA observations of
  the RRL and radio continuum emission at 33~GHz from NGC 253, a
  nearby nuclear starburst galaxy. We detect the previously unobserved
  \hrrl{58} and \hrrl{59} RRLs and make simultaneous sensitive
  measurements of the continuum. We measure integrated line fluxes of
  $44.3 \pm 0.7$~W~m$^{-2}$ and $39.9 \pm 0.8$~W~m$^{-2}$ for the
  \hrrl{58} and \hrrl{59} lines, respectively. The thermal gas in NGC
  253 is kinematically complex with multiple velocity components.  We
  constrain the density of the thermal gas to $1.4 - 4 \times
  10^4$~cm$^{-3}$ and estimate an ionizing photon flux of $1 \times
  10^{53}$~s$^{-1}$. We use the RRL kinematics and the derived
  ionizing photon flux to show that the nuclear region of NGC 253 is
  not gravitationally bound, which is consistent with the outflow of
  gas inferred from the X-ray and \ha\ measurements.  The line
  profiles, fluxes, and kinematics of the \hrrl{58} and \hrrl{59}
  lines agree with those of RRLs at different frequencies confirming
  the accuracy of the previous, more difficult, high frequency
  observations. We find that the EVLA is an order of magnitude more
  efficient for extragalactic RRL observations than the VLA. These
  observations demonstrate both the power of the EVLA and the future
  potential of extragalactic RRL studies with the EVLA.

\end{abstract}


%%----------------------------------------------------------------------
% Main Text
%%----------------------------------------------------------------------

\section{Introduction}

How stars form is one of the most important physical questions in
astrophysics today. While we have viable theories that explain how
star formation occurs in the Milky Way
\citep{2007ARA&A..45..481Z,2007ARA&A..45..565M}, we do not yet
understand how stars form in galaxies with different conditions than
those in the Milky Way. A key piece of the puzzle is quantitative
information about the physical conditions in the interstellar medium
(ISM) in extragalactic star-forming regions with a wide range of
metallicities, total masses, and star formation rates.  Studies of
young star-forming regions still embedded in their natal clouds of
dust and dense gas are particularly important; these embedded regions
give us crucial information about the ISM before the newly-formed
stars have destroyed their birth environments and thus tell us about
the conditions in which stars form.  However, these regions are poorly
understood because they are heavily obscured in the optical and
near-infrared ($A_V \ga 10$), which limits the utility of ``standard''
optical/IR nebular diagnostic lines.

Radio recombination lines (RRLs) are powerful tracers of the physical
properties of obscured thermal gas. These lines occur when electrons
make transitions between high levels ($n \gtrsim 40$) of an excited
atom \citep{1992A&ARv...4..161R,2009ASSL..282.....G}. For typical
extragalactic spatial resolutions, RRLs act like density filters with
higher frequency lines tracing higher density gas
\citep{1997ApJ...482..186Z}. Observations of RRLs at multiple, widely
spaced frequencies constrain the density and filling factor of the
thermal gas in a star-forming region {\em without being affected by
the significant extinction from the dust and gas within the
star-forming region}.  RRLs also provide kinematic information about
the thermal gas and estimates of the ionizing photon flux, the star
formation rate, and the star formation efficiency.  Unfortunately, the
use of RRLs outside the Milky Way system has been limited to a dozen
bright, nearby galaxies because these lines are faint (a few percent
above the continuum) and have large velocity widths
($\gtrsim$~100~\kms). Moreover, the higher frequency RRLs, which trace
the higher density thermal gas and thus younger star forming regions,
have been particularly challenging to observe because the bandwidths
needed to observed these lines were larger than the maximum available
bandwidth of the previous generation of instrumentation.

Work by Anantharamaiah and collaborators established the Very Large
Array (VLA) as a powerful instrument for extragalactic RRL
observations
\citep[e.g.,][]{1993ApJ...419..585A,1996ApJ...472...54Z,1996ApJ...466L..13A,1997ApJ...482..186Z,1998MNRAS.295..156P,2000ApJ...537..613A,2001ApJ...549..896D,2001ApJ...557..659M,2002ApJ...574..701M,2005A&A...435..831R,2005A&A...432....1M,2008A&A...483...79R}.\footnote{\citet{2005A&A...435..831R,2008A&A...483...79R}
  also include data from the Australia Telescope Compact Array.} Most
of these observations were at frequencies $\lesssim$ 8.3~GHz; this was
the highest frequency for which the VLA correlator had a large enough
velocity coverage to observe the entire line using a single correlator
tuning. Observations of RRLs at higher frequencies (43~GHz) by
Rodr{\'{\i}}guez-Rico and collaborators required combining four
correlator tunings in post-processing for the necessary velocity
coverage
\citep{2004ApJ...616..783R,2005ApJ...633..198R,2006ApJ...644..914R,2007ApJ...670..295R,2007ApJ...668..625R}. While
those observations provided vital information on the high density gas
in several systems, they may introduce systematic uncertainties in the
line profile and in the continuum subtraction, particularly for very
broad lines.

The bandwidth of the National Radio Astronomy Observatory (NRAO)
Expanded Very Large Array (EVLA;
\citealp{perley_EVLA_2011})\footnote{The National Radio Astronomy
  Observatory is a facility of the National Science Foundation
  operated under cooperative agreement by Associated Universities,
  Inc.  } WIDAR correlator and the improved sensitivity and additional
frequency coverage of the EVLA receivers are opening new frontiers in
RRL studies. These improvements will make observations of these lines
significantly more efficient than with the VLA enabling us to broaden
the sample of galaxies with RRL measurements. We present observations
taken as part of the EVLA Resident Shared Risk Observing (RSRO)
program AK726 to test the RRL capabilities of the EVLA. The
electronics of the EVLA are completely new, so the performance of the
EVLA for detecting these lines cannot be directly extrapolated from
the very successful attempts using the VLA.

For our pilot study, we chose NGC 253, a nearby galaxy with a nuclear
starburst
\citep{1997ApJ...488..621U,2001ApJ...559..864J,2009MNRAS.392L..16F}. This
galaxy is an ideal source to test the RRL capabilities of the EVLA
because of its strong, well-studied RRL emission; it has measurements
of 11 different RRLs from \hrrl{166} to \hrrl{40}
\citep{1977A&A....60L...1S,1980A&A....82..272M,1990ASSL..163..267A,1996ApJ...466L..13A,1997ApJ...485..143P,2002ApJ...574..701M,2005A&A...432....1M,2006ApJ...644..914R}. We
present observations of two, previously undetected RRLs (\hrrl{58} and
\hrrl{59}). We demonstrate that the EVLA produces results comparable
to previous VLA observations in an eighth of the time and allows
simultaneous observations of multiple RRLs. With the bandwidth of the
EVLA WIDAR correlator, we are able to more accurately derive the line
profile and make excellent, simultaneous measurements of the
continuum.  We use the \hrrl{58} and \hrrl{59} lines to constrain the
density of the thermal gas in NGC 253, estimate the ionizing photon
flux, and explore the kinematics of the thermal gas in the nuclear
region.

We assume a distance to NGC 253 of 3.4 Mpc (1\arcsec\ is 16 pc), which
is the mean of the distances determined from three different fields
using the tip of the red giant branch method
\citep{2009ApJS..183...67D}.

\section{Data Reduction} \label{sec:data-reduction}

The data were taken with the EVLA in DnC configuration on 2010
September 22, 25-30.  The angular resolution was approximately
1.8\arcsec\ with a field of view of 1.4\arcmin. The resolution
corresponds to a physical resolution of 30~pc, which is five to ten
times the size of a typical massive star cluster. While these
observations cannot resolve individual clusters, they are also
unlikely to resolve out diffuse emission associated with the clusters.

We observed \hrrl{58} (32.85220~GHz) and \hrrl{59} (31.22332~GHz),
which were previously undetected in NGC 253. We configured the
correlator to provide two basebands, each containing eight adjacent
128~MHz wide sub-bands with 128 channels and dual polarization. We
tuned the basebands to the sky frequencies corresponding to \hrrl{58}
and \hrrl{59} for NGC 253.  The sensitivity at the edges of the
sub-bands decreases, which increases the noise in this region of the
spectrum by a factor of $\sim 1.4$. The placement of the line peak at
the edge of a sub-band results in the noise being higher in the region
with the line peak. However, this placement allows very accurate
measurements of the flux in the line wings.

We used J0137+3309 (3C48) and J0542+4951 (3C147) as absolute flux
density calibrators, J0319+4130 as the bandpass calibrator, and
J0137-2430 as the complex gain calibrator.  The total on-source time
was 57 minutes.

We used CASA\footnote{\url{http://casa.nrao.edu}} to calibrate the
data using the standard EVLA high frequency calibration
method\footnote{\url{http://casaguides.nrao.edu}} with three
modifications. First, since at the time these data were calibrated
there were no Ka-band flux density calibrator models, we used the
K-band flux density calibrator models. The next two modifications were
used to ensure that each individual sub-band was on the same flux
density scale.  The average opacity value for the central two spectral
windows was used as the opacity value for all spectral windows in a
baseband. Since the opacity values typically vary by only 2-3\% across
the entire baseband, this does not introduce any significant
systematic effects.  Finally, since the flux density derived for the
complex gain calibrator was the same (within the errors) for each
sub-band in one baseband, we set the flux density of the complex gain
calibrator for each sub-band to be the average of the flux densities
determined for the complex gain calibrator.

We self-calibrated using the continuum emission deriving one solution
for each baseband. We continuum subtracted the data in $u-v$ space and
combined the line data (shifting to the same reference frequency) and
the continuum data.  Table \ref{tab:final_image_summary} lists the
image properties.

\section{Results} \label{sec:ngc-253}

\begin{figure}
\centering
\includegraphics[scale=0.5]{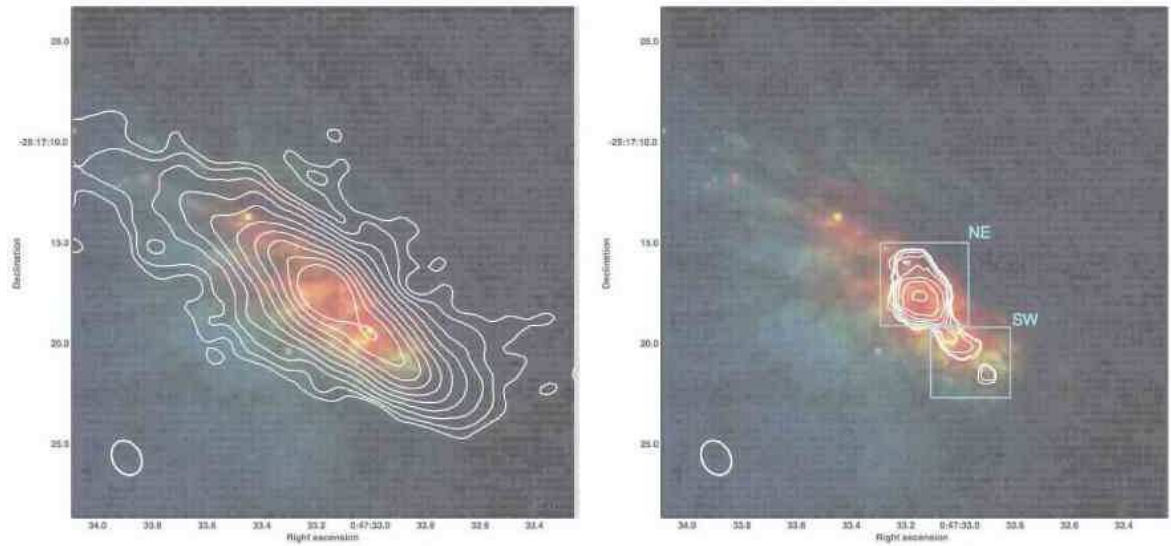}
\caption{{\em HST} ACS (F475W and F814W) and NICMOS (F187N) images of
  NGC 253. The F187N filter (Paschen $\alpha$) is red, the F814W
  filter (I band) is green, and the F475W filter (B band) is
  blue. {\em Left:} 33~GHz continuum contours from 5 to 500 $\sigma$
  on a logarithmic scale (1$\sigma$ = 0.09 \mJybeam). {\em Right:}
  Integrated \hrrl{58} emission contours on a logarithmic scale
  between 5\% and 90\% of the peak flux of 1.15 \Jybeam~\kms. The
  distribution of the 32~GHz continuum and \hrrl{59} line emission is
  identical. The beam ($\sim 1.8\arcsec$) is shown in the lower left
  hand corner of each panel.}
\label{n253_rrl}
\end{figure}

Accurate measurements of both the RRL and continuum emission are
necessary to model the RRL emission. Previous RRL observations
sacrificed continuum sensitivity to detect and spectrally resolve the
RRL. The bandwidth of the EVLA allows us, for the first time, to make
sensitive, simultaneous measurements of both the RRL and continuum
emission. Figure~\ref{n253_rrl} shows the 33~GHz continuum and
integrated \hrrl{58} line flux density for NGC 253 overlaid on an {\em
HST} image of the optical and IR emission. We do not show the 32~GHz
or the \hrrl{59} data since they are very similar.  The strongest RRL
emission is coincident with the peak of the radio continuum
emission. There is another extended region of RRL emission toward the
southwestern portion of the disk. Following
\citet{2006ApJ...644..914R}, we refer to these regions as the
northeastern (NE) and southwestern (SW) regions. We measured the flux
in the NE region using a rectangle with a bottom left corner of
($00^{\rm h}47^{\rm m}33\fs3, -25\degr17\arcmin19$) and a top right
corner of ($00^{\rm h}47^{\rm m}33\fs0, -25\degr17\arcmin15$) and the
SW region using a rectangle with a bottom left corner of ($00^{\rm
h}47^{\rm m}33\fs1, -25\degr17\arcmin23$) and a top right corner of
($00^{\rm h}47^{\rm m}32\fs8, -25\degr17\arcmin19$).

The extinction in this region is extremely high. In the {\em HST}
image, blue (B-band) represents unobscured young star formation, while
the red (Paschen $\alpha$) traces obscured recent star formation. The
RRL emission is coincident with the highly extincted recent star
formation south of the dust lane. The higher resolution of the {\em
  HST} data allows us to distinguish many of the individual star
formation regions that are not resolved in the RRL observations. Even
at the larger spatial resolution of the EVLA data, we can distinguish
two separate, obscured star-forming regions associated with primary
and secondary radio continuum and RRL peaks.

The distribution of the \hrrl{58} line and 33 GHz continuum emission
are similar to the distribution of the \hrrl{53} line and 43~GHz
continuum determined by \citet{2006ApJ...644..914R}. Both data sets
have similar noise levels. The observing times required to reach these
sensitivities, however, are drastically different. The on-source
observing time for the data presented in \citet{2006ApJ...644..914R}
was 8 hours, while the \hrrl{58} data was obtained in 57 minutes. The
observations presented in \citet{2006ApJ...644..914R} only measured
one line and a small amount of continuum, while we measured two lines
as well as obtaining excellent continuum data at both line
frequencies. This comparison shows that the EVLA is at least an order
of magnitude more efficient than the VLA for extragalactic RRL
observations. The 8~GHz bandwidth of the full correlator will deliver
even higher observing efficiencies by providing simultaneous
observations of at least four RRLs at these frequencies.

\begin{figure}
  \centering
  \includegraphics[scale=0.6,angle=-90]{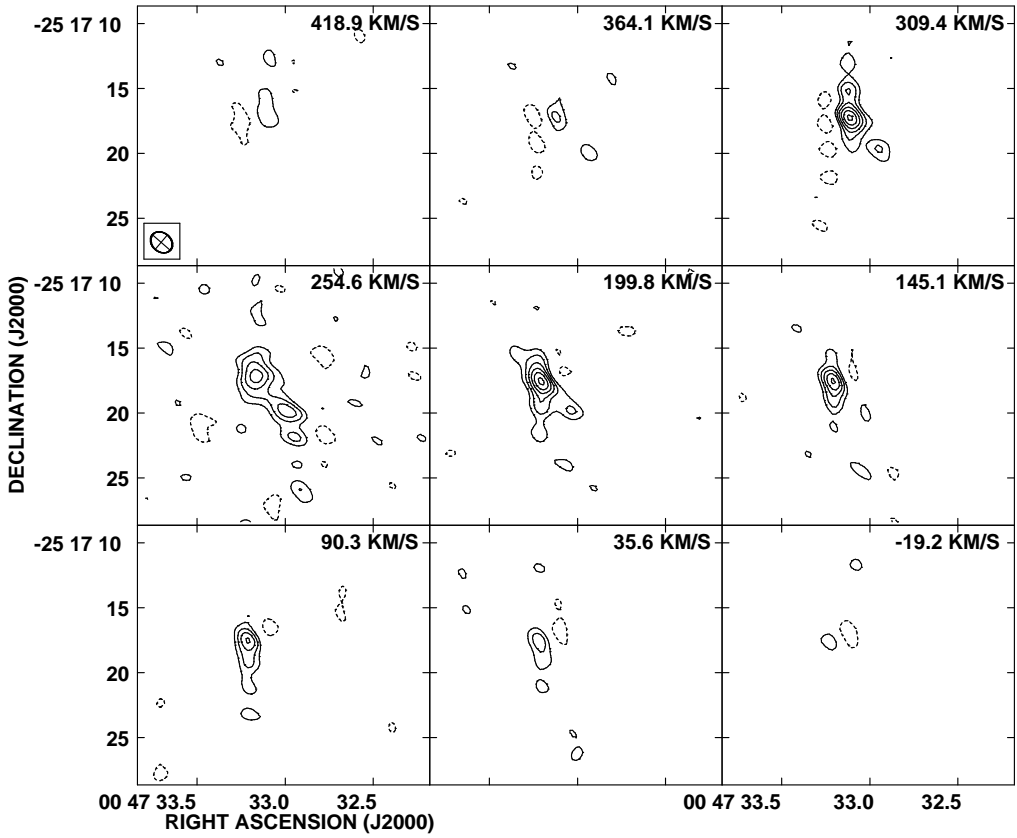}
  \caption{Channel images for \hrrl{58} emission. For this figure, we
    have averaged every three channels together. The contours are
    -3$\sigma$, 3$\sigma$, 5$\sigma$, 7$\sigma$, 9$\sigma$, etc
    ($1\sigma = 0.43 \mJybeam$). The upper right hand corner of each
    panel gives the heliocentric velocity (optical definition). The
    lower left hand corner of the first panel shows the beam ($\sim
    1.8\arcsec$).}
  \label{n253_h58a_chanmap}
\end{figure}

Figure \ref{n253_h58a_chanmap} shows channel maps for the \hrrl{58}
emission. The NE and SW regions have different kinematics. The NE
region covers a wide range of velocities from $\sim 0~\kms$ to
360~\kms\ to, while the SW region covers a much narrower range in
velocities (190~\kms\ to 300~\kms). The central velocities of the NE
and SW region are offset by approximately 60~\kms.
\citet{1996ApJ...466L..13A} attribute the complex \hrrl{92} velocity
field to either a tumbling bar potential, the merger or accretion of
another galaxy, or a warp in the disk in the nuclear
region. \citet{2006ApJ...644..914R} suggest that the RRL velocity
structure could also be the result of a outflow driven by the nuclear
starburst. There is additional X-ray and \ha\ evidence
\citep{2000AJ....120.2965S,2011arXiv1103.1775W} supporting a nuclear
outflow.

\begin{figure}
  \centering
 \includegraphics[scale=0.8]{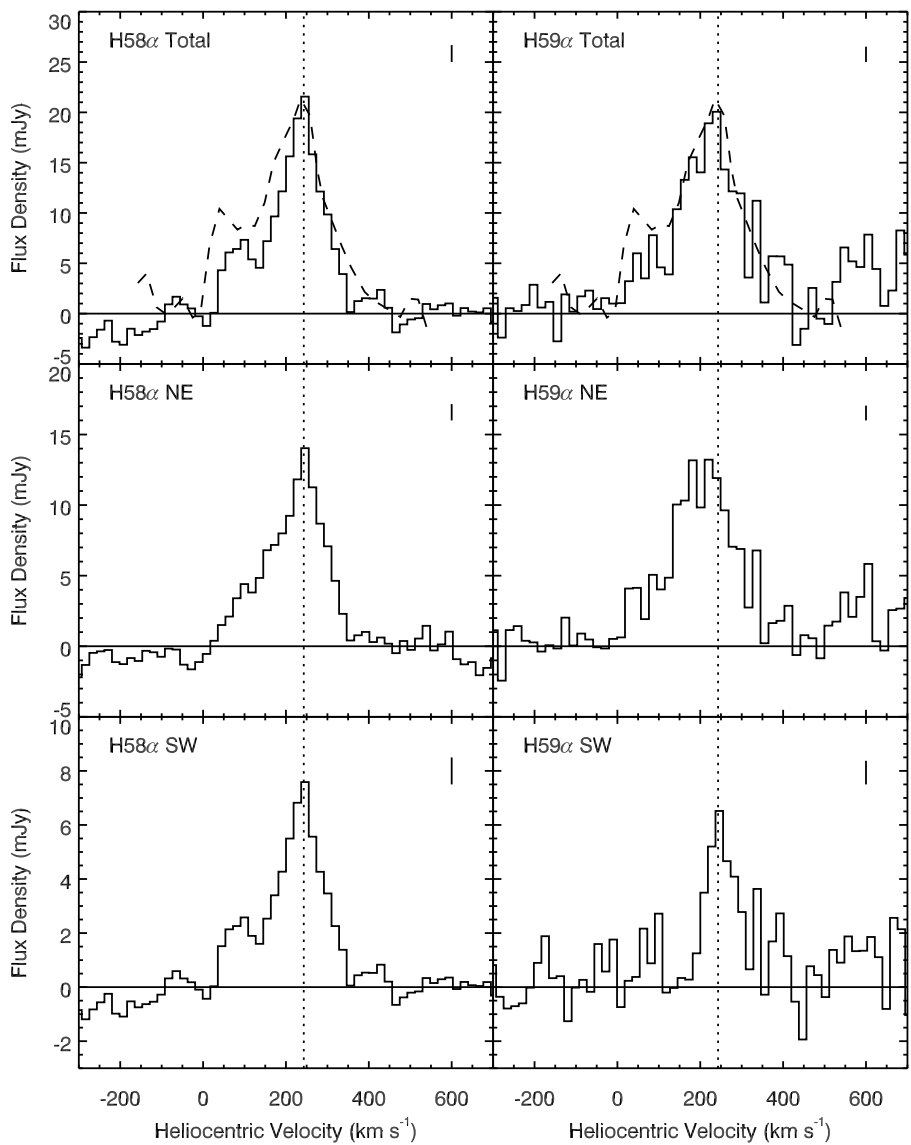}
  \caption{The \hrrl{58} (left panels) and \hrrl{59} (right panels)
    emission for the entire RRL region (top row), the NE region (middle
    row), and the SW region (bottom row). The positions of each region
    are indicated in Figure~\ref{n253_rrl}. The horizontal line
    indicates a flux density of 0~mJy. The dotted vertical line
    indicates the boundary between two sub-bands. The line in the upper
    right hand corner of each panel is the $3\sigma$ per channel error
    for each region. The dashed line in the top row is the \hrrl{53}
    line profile from \citet{2006ApJ...644..914R}. The velocities are
    heliocentric (optical definition). }
  \label{n253_spectra}
\end{figure}

Figure~\ref{n253_spectra} shows the spectrum for the entire line
emitting region and for the NE and SW regions. As expected from the
models of these lines, the higher frequency (\hrrl{58}) line has a
higher peak flux density than the lower frequency line (\hrrl{59}).
The \hrrl{58} and \hrrl{59} line profiles for the entire RRL region
are similar to the line profiles reported by
\citet{2006ApJ...644..914R} and show the presence of several kinematic
components discussed in detail by \citet{2005A&A...432....1M}. We
averaged every three channels in the \hrrl{58} cube together to
compare the peak of the \hrrl{58} line profile to the peak of the
\hrrl{53} line profile from \citet{2006ApJ...644..914R}. The line
profiles agree well.  The \hrrl{53} peak line flux density is larger
than the \hrrl{58} peak line flux density, again in agreement with
phenomenological expectations.

We fit Gaussians to the \hrrl{58} and \hrrl{59} line profiles. While
Gaussians are not the best fit for the data because of the multiple
velocity components, these fits allow us to directly compare our
results to the fits for other RRLs in the
literature. Table~\ref{tab:rrl_parameters} gives the parameters for
the fit to the lines as well as the fits presented in
\citet{2006ApJ...644..914R}. The \hrrl{58} and \hrrl{59} measurements
in this paper agree with the measurements of the line flux from the
literature.  Differences between the flux measurements for each region
in the current and archival data can be attributed to differences in
the regions selected. The differing velocity widths of the \hrrl{58}
and \hrrl{59} are due to systematic uncertainties in the Gaussian fits
because of multiple velocity components and the different
sensitivities of the two images.  The signal to noise of the \hrrl{58}
data is higher than that for the \hrrl{59} data because the \hrrl{59}
line is fainter than the \hrrl{58} line and the noise in the \hrrl{59}
cube is slightly higher.

Extragalactic RRL emission is modeled as a collection of HII regions,
each with an identical electron density ($n_e$), size ($l$), and
temperature ($T$), immersed in a region of diffuse radio continuum
emission
\citep{1991MNRAS.248..585P,1993ApJ...419..585A,2000ApJ...537..613A}. We
use the method of \citet{1997ApJ...482..186Z} to estimate the range of
densities consistent with the \hrrl{58} line data to compare with
those derived in the literature using other RRLs.  Assuming a typical
non-thermal spectral index (-0.7) and a temperature of $10^4$~K,
\hrrl{58} line flux and continuum flux density are consistent with
electron densities of $1.4 - 4 \times 10^4 \, {\rm cm^{-3}}$. This
corresponds to an ionizing photon flux of $10 \times
10^{52}$~s$^{-1}$. Using Equation (1) in \citet{2000ApJ...537..613A},
we obtain star formation rates averaged over the lifetime of OB stars
of 2~${\rm M_\odot \, yr^{-1}}$ in the RRL emission region. The
electron density, ionizing photon fluxes, and star formation rate
estimates derived from the \hrrl{58} line agree with those in the
literature \citep{2005A&A...432....1M,2006ApJ...644..914R}. Our
ionizing photon flux is slightly higher $10 \times 10^{52} \, {\rm
s^{-1}}$ than the $3-6 \times 10^{52} \, {\rm s^{-1}}$ ionizing photon
flux derived by \citet{2005A&A...432....1M}.

We use the derived ionizing photon flux and the measured line widths
to estimate whether the nucleus of NGC 253 is gravitationally bound
\citep[e.g.][]{2003Natur.423..621T}. We estimate the mass of the
cluster from the number of ionizing photons using
\citet{1999ApJS..123....3L}. We assume that the cluster was formed in
an instantaneous burst, is 3~Myr old \citep{2009MNRAS.392L..16F}, and
has a Salpeter IMF with an upper mass limit of 100~$M_\odot$, which
gives a cluster mass of $4.0 \times 10^6$~M$_\odot$. The mass of the
ionized gas derived by \citet{2006ApJ...644..914R} is negligible
compared to the cluster mass.  The line emitting region is 53 pc by
135 pc. The escape velocity for the nucleus is then between 17 \kms\
and 26 \kms. This width is much smaller than the velocity width of the
RRL emission. We assume that the RRL line widths are dominated by the
motion of the gas since they are much larger than the intrinsic width
of a RRL from a single HII region ($\sim 20 \, \kms$). The escape
velocity is still much smaller the RRL velocity width if we consider
only the NE region. There we derive an ionizing photon flux of $9.3
\times 10^{52} \, {\rm s^{-1}}$, a cluster mass of $3.97 \times
10^6$~M$_\odot$ and an escape velocity of 25~\kms. These calculations
show that the central region of NGC 253 is most likely not
gravitationally bound, supporting the hypothesis that there is an
outflow of gas from the nucleus of this galaxy.

Although RRLs provide an extinction-free tracer of the gas ionized by
the cluster, the use of RRL data may underestimate the number of
ionizing photons if the photons are absorbed by dust rather than
ionize hydrogen. Therefore, the mass of the cluster is a lower
limit. To obtain an escape speed comparable to the RRL velocity width,
the number of ionizing photons would have to be underestimated by a
factor of 45.

\section{Conclusions} \label{sec:concl-future-work}

We have presented observations of the \hrrl{58} and \hrrl{59} RRLs and
the continuum emission at 32 and 33~GHz from the nucleus of NGC
253. We measure integrated line fluxes of $44.3 \pm 0.7$ and $39.9 \pm
0.8$~W~m$^{-2}$ for the \hrrl{58} and \hrrl{59} lines respectively. We
constrain the thermal gas density to $1.4 - 4 \times
10^4$~cm$^{-3}$. We derive an ionizing photon flux of $10 \times
10^{52}$~s$^{-1}$.  These estimates of the density and ionizing photon
flux are consistent with estimates from the literature
\citep{2002ApJ...574..701M,2005A&A...432....1M,2006ApJ...644..914R}.

The kinematics of the thermal gas in NGC 253 are complex. The NE
region has a broader velocity width (170 -- 200 \kms) than the SW
region (145 -- 159 \kms). From our estimate of the ionizing photon
flux, we estimate a cluster mass for the nuclear region of NGC 253 of
$4 \times 10^6$~M$_\odot$. The escape velocity for this mass is only
17 \kms\ to 26 \kms: much lower than the \hrrl{58} and \hrrl{59}
velocity widths. Therefore, the nuclear region of NGC 253 is most
likely not gravitationally bound. We suggest that an outflow of gas
from the nuclear region is the most likely explanation for the thermal
gas kinematics of NGC 253. The presence of this outflow has been
inferred from X-ray \citep{2000AJ....120.2965S} and \ha\
\citep{2011arXiv1103.1775W} observations.

In contrast to previous VLA high frequency RRL observations, we were
able to detect not one, but two RRLs, and make much more sensitive
measurements of the continuum in an eighth of the time on
source. Accurate measurements of both the RRL and continuum emission
are necessary to model the RRL emission from these regions in
detail. The EVLA is currently an order of magnitude more efficient for
high frequency RRL observations than the VLA. The EVLA observing
efficiency will increase even further once the full 8~GHz bandwidth is
available.

These observations demonstrate the potential of the EVLA to
revolutionize extragalactic RRL observations. Future EVLA observations
will be able to use RRLs to measure the properties of the thermal gas
in a wider variety of galaxies. Ultimately, these observations will
lead to a quantitative picture of how the properties of the ISM affect
the resulting stellar populations.

\acknowledgments The authors acknowledge the hard work and dedication
of the EVLA commissioning team. K.E.J. acknowledges support provided
by NSF through CAREER award 0548103 and the David and Lucile Packard
Foundation through a Packard Fellowship.

{\it Facilities:} \facility{EVLA}

%%----------------------------------------------------------------------
%%                            Appendices
%%----------------------------------------------------------------------

%%----------------------------------------------------------------------
%%                            Bibliography
%%----------------------------------------------------------------------

%%----------------------------------------------------------------------
%%                              Tables
%%----------------------------------------------------------------------

\begin{deluxetable}{lcccccccc}
\tablewidth{0pt}
\tabletypesize{\scriptsize}
\tablecaption{Image Properties \label{tab:final_image_summary}}
\tablehead{
\colhead{} &
\colhead{Frequency} &
\colhead{} &
\colhead{} &
\colhead{Beam} &
\colhead{PA} &
\colhead{Noise} &
\colhead{Velocity Resolution} &
\colhead{Bandwidth}\\
\colhead{Line/IF} &
\colhead{GHz} &
\colhead{Type} &
\colhead{Robust} &
\colhead{\arcsec\ } &
\colhead{\degr} &
\colhead{\mJybeam} &
\colhead{\kms} &
\colhead{MHz}}
\startdata
\hrrl{58}   &  32.85220      &   Line         &  0       &  $1.82 \times 1.46$  &  49.1     & 0.75       &  18.3     & \nodata \\
\hrrl{59}   &  31.22331      &   Line         &  0       &  $1.80 \times 1.62$  &  60.3     & 1.3        & 19.2      & \nodata \\ \hline
 IF1        &  33.0          &   Continuum    &  0       &  $1.75 \times 1.38$  &  58.5     & 0.09       & \nodata   & 768   \\  
 IF2        &  32.0          &   Continuum    &  0       &  $1.85 \times 1.40$  &  64.8     & 0.16       & \nodata   & 768   \\  
\enddata
\end{deluxetable}

\begin{deluxetable}{lllllllll}
\tablewidth{0pt}
\tablecolumns{9}
\tabletypesize{\scriptsize}
\tablecaption{ RRL Parameters \label{tab:rrl_parameters}}
\tablehead{
\colhead{} &
\colhead{}   &
\colhead{$\nu_{rest}$}    & 
\colhead{$S_c$}  &
\colhead{$S_L$}  &
\colhead{$\Delta V_{FWHM}$}  &
\colhead{$V_{hel}$} &
\colhead{$S_L \Delta V_{FWHM}$} &
\colhead{} \\
\colhead{Region} &
\colhead{Line} &
\colhead{GHz} &
\colhead{mJy} &
\colhead{mJy} &
\colhead{\kms} &
\colhead{\kms} &
\colhead{$10^{-22} {\rm W m^{-2}}$} &
\colhead{Ref.} }
\startdata
total    &  \hrrl{58}  & 32.85220 & $314.27 \pm 0.05$   &  $17.5 \pm 0.3$ &  $158 \pm 3$   & $231 \pm 1$   &   $44.3 \pm 0.7$    &  2   \\
total    &  \hrrl{59}  & 31.22331 & $332.5  \pm 0.1$    &  $16.3 \pm 0.2$ &  $209 \pm 3$   & $222 \pm 1$   &   $39.9 \pm 0.8$    &  2   \\ \hline
NE       &  \hrrl{58}  & 32.85220 & $241.63 \pm 0.04$   &  $11.4 \pm 0.2$ &  $169 \pm 3$   & $228 \pm 1$   &   $28.6 \pm 0.5$    &  2   \\
NE       &  \hrrl{59}  & 31.22331 & $260.06 \pm 0.07$   &  $11.8 \pm 0.2$ &  $200 \pm 3$   & $206 \pm 1$   &   $26.8 \pm 0.5$    &  2    \\ \hline
SW       &  \hrrl{58}  & 32.85220 & $83.19  \pm  0.03$  &   $6.1 \pm 0.2$ &  $159 \pm 5$   & $231 \pm 2$   &   $15.6 \pm 0.4$    &  2   \\
SW       &  \hrrl{59}  & 31.22331 & $84.77 \pm 0.05$    &   $4.6 \pm 0.1$ &  $145 \pm 7$   & $256 \pm 3$   &   $12.8 \pm 0.4$    &  2    \\
\cutinhead{Archival Data} 
total    &  \hrrl{53}  & 43.3094  &  $360 \pm 20$ & $21 \pm 2$    & $230 \pm 20$  & $210 \pm 10$ & $69 \pm     9$ &  1   \\
total    &  \hrrl{92}  &  8.3094  &  $605 \pm 10$ & $9.0 \pm 0.5$ & $190 \pm 10$  & $206 \pm 4$  & $4.7 \pm 0.5$  &  1   \\ \hline
NE       &  \hrrl{53}  & 43.3094  &  $134\pm 5$   & $9.0 \pm 0.5$ & $200 \pm 10$  & $220 \pm 5$  & $25 \pm 1$     &  1   \\
NE       &  \hrrl{92}  &  8.3094  &  $278 \pm 10$ & $4.5 \pm 0.1$ & $190 \pm 5$   & $200 \pm 2$  & $2.4 \pm 0.2$  &  1   \\ \hline
SW       &  \hrrl{53}  & 43.3094  &  $24 \pm 2$   & $2.8 \pm 0.3$ & $130 \pm 20$  & $215 \pm 10$ & $4 \pm 1$      &  1   \\
SW       &  \hrrl{92}  &  8.3094  &  $40 \pm 4$   & $0.7 \pm 0.1$ & $160 \pm 15$  & $220 \pm 10$ & $0.3 \pm 0.03$ &  1   \\ \hline
\enddata
\tablerefs{ (1) \citet{2006ApJ...644..914R}; (2) This work }
\end{deluxetable}

%%----------------------------------------------------------------------
%%                              Figures
%%----------------------------------------------------------------------

%%% Figure 1

%%% Figure 2

%%% Figure 3

\clearpage

\end{document}